\documentstyle[epsfig,graphics,graphicx,preprint,aps]{revtex}
\begin{document}
\draft
\title{Formation of Schwarzschild black hole from the 
scalar field collapse in four-dimensions.} 
\author{G. Oliveira-Neto\thanks{Email:
gilneto@fisica.ufjf.br} and F. I. 
Takakura\thanks{Email:takakura@fisica.ufjf.br}} 
\address{Departamento de F\'{\i}sica,
Instituto de Ciencias Exatas,
Universidade Federal de Juiz de Fora,
CEP 36036-330, Juiz de Fora,
Minas Gerais, Brazil.}
\date{\today}
\maketitle

\begin{abstract}
We obtain a new self-similar solution to the Einstein's 
equations in four-dimensions, representing the collapse of a
spherically symmetric, minimally coupled, massless, scalar 
field. Depending on the value of certain parameters, this
solution represents the formation of naked singularities 
and black holes. Since the black holes are identified as 
the Schwarzschild ones, one may naturally see how these black
holes are produced as remnants of the scalar field collapse. 
\end{abstract}

Since the first analytical model of gravitational collapse
proposed, using general relativity \cite{snyder}, many works 
have been done in this area. An important model is the one 
where the collapsing matter is represented by a massless 
scalar field and the space-time has a spherical symmetry 
(Einstein-scalar model). D. Christodoulou has pioneered 
analytical studies of that model \cite{chris}. In particular, 
he showed that for a self-similar scalar field collapse 
there are initial conditions which result in the formation 
of black holes, naked singularities and bouncing solutions. 

Many analytical results obtained by Christodoulou have
been confirmed, by numerical studies of the Einstein-scalar 
model, by M. W. Choptuik \cite{choptuik}. Other authors have 
investigated the Einstein-scalar model numerically and 
analytically confirming the above results \cite{wang0}. In 
particular, we may mention the works in references 
\cite{brady}, \cite{tomimatsu}. There, the authors were 
able to reproduce qualitatively, in an analytical solution 
\cite{roberts}, the results derived numerically in 
\cite{choptuik}.

The Roberts solution \cite{roberts}, has two arbitrary
parameters $\alpha$ and $\beta$. In the work in reference 
\cite{brady}, the author fixed $\beta = 1$ and let $\alpha$ 
vary in the range $[1/4, -\infty]$. Once $\beta$ is fixed, 
$\alpha$ determines also the scalar field strength. It will
be zero for $\alpha = 1/4$ and it will increase for 
decreasing values of $\alpha$. The scalar field eventually 
diverges for $\alpha \to -\infty$. Starting with 
$\alpha = 1/4$, the scalar field is zero and one has 
Minkowski space-time. Then, for $0 < \alpha < 1/4$, although 
the field is non-longer zero its strength is not enough to 
initiate the collapse. The space-time in this case 
represents a bouncing. For $\alpha = 0$ the gravitational
collapse begins and a naked singularity is formed at $r = 0$.
If one increases even further the field strength by letting
$\alpha$ takes values in the range $-\infty < \alpha < 0$, 
one has
the black hole formation. Therefore, if one turns the scalar
field on and gradually increases its strength one will see
all the collapse stages described above. Suppose now, one
decides to turn the scalar field off, which will eventually
happen for one has a limited amount of energy. One will
see all the collapse stages described above, in the reverse
order. Starting with the black hole and finishing with the
Minkowski space-time. So, we may conclude that the Roberts 
solution does not produce an eternal black hole that will be 
left as a remnant of the scalar field collapse. Once the 
scalar field is turned off one loses all trace of the 
collapse. It is not difficult to see that one may obtain
the above result even if one let $\beta$ vary in an 
appropriate range.

One expects that the Schwarzschild black hole results from 
the collapse of a spherical symmetric neutral scalar field. 
So far, the only way to obtain this black hole as the result 
of the scalar field collapse was by joining an analytical 
solution to the Einstein-scalar equations that represents
the collapse with the Vaidya solution \cite{wang}. Of course,
this procedure generates the Schwarzschild black hole since
it is a particular case of the Vaidya one. In reference 
\cite{wang}, the authors joined the Roberts solution 
\cite{roberts} with the exterior Vaidya solution. 

In the present paper we would like to present a new solution 
to the Einstein's equation representing the self-similar, 
spherically symmetric, collapse of a minimally coupled, 
neutral,
massless, scalar field, in four-dimensions. As we shall
see this solution has two parameters, in the same way as
Roberts' solution. Depending on the value of these 
parameters, we may have the formation of naked singularities 
and black holes. Since the black holes are identified as 
the Schwarzschild ones, we shall see how these black holes 
are produced as remnants of the scalar field collapse. 

We shall start by writing down the ansatz for the space-time 
metric. As we have mentioned before, we would like to consider 
the spherically symmetric, self-similar, collapse of a 
massless scalar field in four-dimensions. Therefore, we shall 
write our metric ansatz as,

\begin{equation}
\label{1}
ds^2\, =\, -\, 2 e^{2\sigma(u,v)} du dv\, +\, r^2(u,v) 
d\Omega^2\, ,
\end{equation}
where $\sigma(u,v)$ and $r(u,v)$ are two arbitrary functions 
to be determined by the field equations, $(u,v)$ is a pair of 
null coordinates varying in the range $(-\infty,\infty)$, and 
$d\Omega^2$ is the line element of the unit sphere.

The scalar field $\Phi$ will be a function only of the two 
null coordinates and the expression for its stress-energy 
tensor $T_{\alpha\beta}$ is given by \cite{wheeler},

\begin{equation}
\label{2}
T_{\alpha\beta}\, =\, \Phi,_\alpha \Phi,_\beta\, -\,
{1\over 2} g_{\alpha\beta} \Phi,_\lambda \Phi^{,_\lambda}\, .
\end{equation}
where $,$ denotes partial differentiation.

Now, combining Eqs. (\ref{1}) and (\ref{2}) we may obtain the
Einstein's equations which in the units of Ref. \cite{wheeler}
and after re-scaling the scalar field, so that it absorbs the 
appropriate numerical factor, take the following form,

\begin{equation}
\label{3}
2 ( r r,_{uv}\, +\, r,_u r,_v )\, +\, e^{2\sigma}\, =\, 0\, ,
\end {equation}
\begin{equation}
\label{4}
2 r r,_{vv}\,-\, 4 r r,_v \sigma,_v\, =\, -\, r^2 
(\Phi,_v)^2\, ,
\end{equation}
\begin{equation}
\label{5}
2 r r,_{uu}\,-\, 4 r r,_u \sigma,_u\, =\, -\, r^2 
(\Phi,_u)^2\, ,
\end{equation}
\begin{equation}
\label{6}
2 ( r^2 \sigma,_{uv}\, +\, r r,_{uv} )\, =\, -\, r^2 
( \Phi,_u \Phi,_v )\, ,
\end{equation}
The equation of motion for the scalar field, in these 
coordinates, is

\begin{equation}
\label{7}
r \Phi,_{uv}\, +\, r,_u \Phi,_{v}\, +\, r,_v \Phi,_{u}\,
=\, 0\, .
\end{equation}

The above system of non-linear, second-order, coupled, 
partial differential equations (\ref{3})-(\ref{7}) can be 
solved if we impose that it is continuously self-similar. 
More precisely, the solution assumes the existence of an 
homothetic Killing vector of the form,

\begin{equation}
\label{8}
\xi\, =\, u {\partial \over \partial u}\, +\, v 
{\partial \over \partial v}\, ,
\end{equation}
Following Coley \cite{coley}, equation (\ref{8}) 
characterizes a self-similarity of the first kind. We 
can express the solution in terms of the variable 
$z = v/u$. 

In order to obtain our solution, we shall 
implement the self-similarity in the above equations 
(\ref{3}-\ref{7}) in a slightly different way than
previous works \cite{brady}, \cite{tomimatsu}. We 
shall write the unknown functions as: 
$\Phi(u,v)=\Phi(z)$, 
$\sigma(u,v)=\sigma(z)$ and $r(u,v)=r(z)$. It is
clear that the main difference between our work and 
previous ones is in the way we write $r$. The system
of equations (\ref{3}-\ref{7}) will have the mentioned
self-similarity if we re-write $e^{2\sigma}$ in 
(\ref{1}) in the following way,

\begin{equation}
\label{9}
e^{2\sigma}\, =\, f,_u f,_v\, ,
\end{equation}
where $f = f(u,v)$ is an arbitrary function of the
null coordinates $u$ and $v$. $f$ will also be 
written as $f(z)$ when we implement the 
self-similarity.

In terms of $z$ and taking in account (\ref{9}), the 
system of equations (\ref{3}-\ref{7}) becomes,

\begin{equation}
\label{10}
2 r ( r'\, +\, z r'' )\, +\, 2 z ( r' )^2\, +\, z 
F^2\, =\, 0\, ,
\end{equation}
\begin{equation}
\label{11}
2 z r''\, -\, {2 r'\over F} ( F\, +\, 2 z F' )\, 
=\, -\ z r ( \Phi' )^2\, ,
\end{equation}
\begin{equation}
\label{12}
2 z r''\, +\, 2 r'\, +\, {2 r\over F^2} [ F F'\, 
+\, z F F''\, -\, z ( F' )^2 ) ]\, =\, -\ z r 
( \Phi' )^2\, ,
\end{equation}
\begin{equation}
\label{13}
(z r^2 \Phi' )'\, =\, 0\, ,
\end{equation}
where $'$ means differentiation with respect to $z$,
$f' \equiv F$ and equations (\ref{4}) and (\ref{5}) 
reduce to equation (\ref{11}).

The above system, equations (\ref{10}-\ref{13}), can
be greatly simplified if we re-write it in terms of
the new variable $t \equiv \ln z$. The resulting
equations are,

\begin{equation}
\label{14}
(r^2)^{\cdot \cdot}\, =\, -\, e^{2 t} F^2\, ,
\end{equation}
\begin{equation}
\label{15}
2 \ddot{r}\, -\, 4 \dot{r} \left( 1\, +\, 
{\dot{F}\over F} \right)\, =\, -\, r \dot{\Phi}^2\, ,
\end{equation}
\begin{equation}
\label{16}
2 \ddot{r}\, +\, 2 r \left( {\dot{F}\over F} 
\right)^\cdot\, =\, -\, r \dot{\Phi}^2\, ,
\end{equation}
\begin{equation}
\label{17}
( r^2 \dot{\Phi} )^\cdot\, =\, 0\, ,
\end{equation}
where the $\cdot$ means differentiation with respect
to $t$.

We solve the system of equations (\ref{14}-\ref{17}) 
by initially writing down a second order differential 
equation for $r$ in the following way. First of all,
we subtract equation (\ref{15}) from equation 
(\ref{16}) and manipulate it in order to find,

\begin{equation}
\label{18}
1\, +\, {\dot{F}\over F}\, =\, {C\over r^2}\, ,
\end{equation}
where $C$ is a real integration constant. Later, for 
a particular situation, we shall be able to identify 
the physical meaning of $C$. Then, we differentiate
equation (\ref{14}) once with respect to $t$ and 
introduce, in the differentiated equation, the value 
of $- e^{2 t} F^2$ from equation (\ref{14}) and the 
information coming from equation (\ref{18}). Finally, 
we integrate the resulting equation to find,

\begin{equation}
\label{19}
r^3 \ddot{r}\, -\, 2 C r \dot{r}\, =\, B\, ,
\end{equation}
where $B$ is a real integration constant.  

The physical meaning of $B$ may be understood if we
eliminate $F$ and its derivatives from either one of
the equations (\ref{15}) or (\ref{16}), with the aid
of equation (\ref{18}). They give the same result,
which is,

\begin{equation}
\label{20}
\dot{\Phi}\, =\, {\sqrt{-2B}\over r^2}\, .
\end{equation}
It is easy to see that equation (\ref{17}) also leads
to the same result above.

Observing equation (\ref{20}), we learn that $B$ is
a constant associated with the scalar field strength.
More than that, we also learn that it cannot take 
positive values. Its domain is restricted to 
$[0,-\infty]$. It is important to notice that equation
(\ref{20}) gives us a way to compute the scalar field
once we know $r$ as a function of $t$.

In order to understand the meaning of the real 
constant $C$ in equation (\ref{18}), we shall consider
the simplified situation where $B=0$. From equation
(\ref{20}) we see that $\Phi$ is a constant. Without
lost of generality we may set this constant to zero.
It means that we are now dealing with a spherically
symmetric vacuum solution to the Einstein's equations,
so from Birkhoff's theorem that solution is 
necessarily a piece of the Schwarzschild space-time
\cite{birkhoff}.

We may identify the role played by $C$ in the
Schwarzschild space-time by first solving equation
(\ref{19}) for the present situation. This gives,

\begin{equation}
\label{21}
t\, =\, {r\over D}\, +\, {2 C\over D^2} \ln ( D r\,
-\, 2 C )\, +\, E\, ,
\end{equation}
where $D$ and $E$ are integration constants. Now, using 
the general expression of $e^{2 \sigma}$, which is derived 
from equation (\ref{9}) and has the following value,

\begin{equation}
\label{22}
e^{2 \sigma}\, =\, {- e^{2 t} F^2\over u v}\, ,
\end{equation}
we may compute $g_{uv}$ for our present situation.
With the aid of equations (\ref{14}) and (\ref{21}),
we get,

\begin{equation}
\label{23}
e^{2 \sigma}\, =\, {2 D ( D r\, -\, 2 C )\over 
u v r}\, .
\end{equation}
Finally, we may evaluate the coordinate independent
quantity $g^{\alpha \beta} r,_{\alpha} r,_{\beta}$,
for the present situation, and compare it with the
same quantity evaluated for the Schwarzschild 
space-time. Using the values of $r$ from equation 
(\ref{21}) and $g^{uv}$ easily derived from equation 
(\ref{23}), we have,

\begin{equation}
\label{24}
g^{\alpha \beta} r,_\alpha r,_\beta\, =\, {( D r\, 
-\, 2 C )\over D r}\, .
\end{equation}
Since the same quantity for the Schwarzschild 
space-time is $1 - 2 M/r$, where $M$ is the
Schwarzschild mass, we conclude that if we set
$D = 1$, $C$ may be identified as the Schwarzschild 
mass. It is important to notice that this
identification also tell us that $C$ should not be 
negative in the general case ($B, C \neq 0$). If it 
were negative we would not be able to recover the 
Schwarzschild space-time as a limiting case of the 
general case. It is important to mention at this
point that if one sets $C=0$ and $B=0$ one gets
Minkowski space-time.

With the aid of equations (\ref{21}) and (\ref{23}),
for $D=1$ and $C=M$, we may write the line element 
equation (\ref{1}) for the Schwarzschild solution,

\begin{equation}
\label{24,1}
ds^2\, =\, -\, {4\over uv}\left(r\, -\, 
{2M\over r}\right) du dv\, +\, r^2(u,v) 
d\Omega^2\, .
\end{equation}
One may transform this line element equation 
(\ref{24,1}) to the usual one in terms of the 
($\bar{t}, r$) coordinates,

\begin{equation}
\label{24,2}
ds^2\, =\, -\, \left(1\, -\, {2M\over r}\right) 
d\bar{t}^2\, +\, {dr^2\over \left(1\, -\, 
{2M\over r}\right)}\, +\, r^2 d\Omega^2\, .
\end{equation}
In order to do that one has to use equation 
({\ref{21}) and the coordinate transformation
$\bar{t} = \ln (uv)$.

Let us consider the general case ($B, M \neq 0$). 
Equation (\ref{19}) cannot be solved analytically,
as far as we know, therefore we solve it numerically.
This equation can be written in the following
suggestive form,

\begin{equation}
\label{25}
\left( \dot{r}\, +\, {2 M\over r} \right)^\cdot\, =\,
{B\over r^3}\, .
\end{equation}
Equation (\ref{25}) reduces to the equation governing
the case $B=0$ when we set $r \to \infty$. This means
that $r(t)$ in the general case behaves asymptotically 
in the same manner as $r(t)$ in the Schwarzschild 
space-time. Therefore, the space-times in the general
case are asymptotically flat.

Then, taking in account this result we search for
numerical solutions of equation (\ref{19}) which
asymptotically satisfy equation (\ref{21}) and its
derivative with respect to $t$, both with $D=1$.
From Eq. (\ref{21}) we obtain that in the limit 
$r \to \infty$ we have $r(t) = t$ and $\dot{r}(t) = 1$.
We have noticed that for a large number of different
asymptotic values of $r$ one obtains the same 
qualitative behavior of the solutions. In Figure $1$,
we exemplify this fact by tracing two graphics $r 
\times t$ for the space-time with $M=1$ and $B=-40$, 
which we shall call $ST140$. In the first one, Figure 
$1(a)$, we set the asymptotic value of $r$ to 
$10^6$. Therefore, the asymptotic conditions to 
solve Eq. (\ref{19}) are $r(10^6) = 10^6$ and 
$\dot{r}(10^6) = 1$. In Figure $1(b)$, we set the 
asymptotic value of $r$ to $10^3$. As a matter of 
definitiveness we have chosen in our calculations 
the same set of asymptotic conditions of Figure 
$1(a)$.

For real, positive values of $M$ and real, negative
values of $B$ we find an infinity set of solutions
satisfying the above conditions. We shall call this 
set $S$. The $r(t)$'s, of the space-times in $S$, 
have the following behavior. They start collapsing 
from $+\infty$ initially linearly with $t$ and then 
deviating from this regime until they reach $r=0$, 
for certain values of $t$. These values of $t$ depend 
in an unknown manner on $B$ and $M$. In Figure $2(a)$, 
we can see $r$ as a function of $t$ for the space-time 
$ST140$ and in Figure $2(b)$ for the space-time with 
$M=3$ and $B=-70$, which we shall call $ST370$.

From the curvature scalar $R$ for the space-times
in $S$, which is given by,

\begin{equation}
\label{26}
R\, =\, {- 2 B\over r^2 ( r^2 \dot{r}^2\, +\, 2 M r
\dot{r}\, +\, B )}\; ,
\end{equation}
we can see that $r=0$ is a physical singularity.
Performing a numerical investigation of a sufficiently
large number of space-times in $S$ we conclude that
$r=0$ is the only physical singularity of these
space-times. In Figure $3(a)$, we can see how $R$ 
behaves as a function of $t$ for the space-time 
$ST140$ and in Figure $3(b)$ for the space-time 
$ST370$

Now, we may look for the presence of an apparent
horizon in the solutions in $S$. We do that
by investigating any change of sign in the quantity, 
$g^{\alpha \beta} r,_\alpha r,_\beta$. This quantity 
has the below value for the space-times in $S$,

\begin{equation}
\label{27}
g^{\alpha \beta} r,_\alpha r,_\beta\, =\, {r^2 
\dot{r}^2\over r^2 \dot{r}^2\, +\, 2 M r
\dot{r}\, +\, B}\, .
\end{equation}
Performing a numerical investigation of a sufficiently
large number of space-times in $S$ we conclude that
they do not develop an apparent horizon. For large
values of $t$, $g^{\alpha \beta} r,_\alpha r,_\beta$ 
equation (\ref{27}) is a positive constant which 
has the value $1$. As $t$ decreases, it 
grows positive until it blows up at the $r=0$ 
singularity. Figure $4(a)$ shows the curve $g^{\alpha
\beta} r,_\alpha r,_\beta$ versus $t$ for the
space-time $ST140$ and Figure $4(b)$ for space-time 
$ST370$. We may also conclude, from the above result, 
that $r=0$ is a time-like singularity.

Finally, we solve equation (\ref{20}) for the 
space-times in $S$, together with the condition that 
$\Phi$ asymptotically vanishes. Using the same 
asymptotic values of $r$ and $t$, mentioned 
above, this condition translates to $\Phi(10^6) = 0$.
Performing a numerical investigation of a 
sufficiently large number of space-times in $S$ we 
find a scalar field with the following behavior.
It grows negative from zero at $t \to \infty$, 
initially very slowly, then more quickly as we 
diminish $t$. As we approach the physical singularity
$\Phi$ grows negative very quickly and finally blows 
up at the singularity. Figure $5(a)$ shows $\Phi$ as a 
function of $t$ for the space-time $ST140$ and Figure 
$5(b)$ for the space-time $ST370$. It is not difficult
to verify numerically that, in the asymptotic region, 
$\Phi$ is $C^N$ and its derivatives with respect to $t$ 
vanish as $t \to \infty$.

As a matter of completeness we mention that the 
particular case $M=0$, $B<0$, has an analytical solution 
with the same qualitative behavior of the numerical 
solution of the general case ($M>0$, $B<0$). The main 
difference is that in the first case if one takes $B=0$ 
one obtains flat space-time instead of Schwarzschild 
space-time. Also, it is not difficult to see numerically
that the so-called `bouncing solutions', present in the
collapse solutions of References \cite{chris}, 
\cite{choptuik}, \cite{brady}, \cite{tomimatsu}, can 
only appear from equation ({\ref{19}) for the unphysical 
situation of $M > 0$, $B >0$.

Collecting together all the above results we conclude
that for the infinite number of space-times in $S$, 
parametrized by the constants $M$ ($M > 0$) and $B$
($B < 0$), $\Phi$ will collapse to $r=0$ without the 
formation of an apparent horizon. Since $r=0$ is a 
physical singularity for all elements of $S$, this 
collapse will give rise to naked singularities. These
singularities are time-like and the space-times are
asymptotically flat. Combining together the above 
properties of these naked singularities with the fact 
that they define a region of non-zero measure in the 
parameter space of solutions, we can say that they 
qualify as possible counter-examples to the weak 
\cite{penrose}, \cite{wald} and strong \cite{penrose1} 
cosmic censorship conjectures. An issue that will be 
further investigated elsewhere.

It is important to describe the space-times in $S$
in terms of the original null coordinates ($u , v$).
We start this program by noticing that, since 
$t = \ln{(v/u)}$ we have the following different 
ways to obtain the limit $t \to \infty$ if $(u, v)$
are non-negative. They are: (i) $v \to \infty$ while 
$u$ remains finite and (ii) $u \to 0$ while $v$ is 
non-zero. We may also have the  limit $t \to \infty$ 
if $(u, v)$ are non-positive. Once they are physically 
equivalent to (i) and (ii) above, we shall restrict 
our attention to non-negatives $(u, v)$. Another
important information is the location of $r=0$ in
the ($u , v$) coordinates. As we have seen above,
there is a certain $t$, let's say $t_0$, associated
with the physical singularity. In terms of ($u , v$),
$r=0$ is the straight line passing through the 
origin with the equation: $u = \exp{(-t_0)} v$.
Collecting together the above information, we see
that the space-times in $S$ may be described as
a piece of the $u \geq 0$ and $v \geq 0$ sector in 
$(u,v)$ coordinates. Since, in the surface $u = 0$ 
the space-times are flat, we may extend them, 
smoothly, to the region $u < 0$ as Minkowski 
space-times. Figure $6$ shows the conformal diagram 
for a typical space-time in $S$. We may see that the 
space-time is divided, naturally, in two distinct 
regions. The first one is the Minkowskian region 
where $u < 0$ (I). Then, we have the collapse region 
where $v >0$ and $u > 0$ (II). We can interpret
this diagram in the following way: the scalar field 
starts collapsing from $u = 0$, the initial data 
surface in the present coordinates. The space-time
is flat and $\Phi$ and its derivatives are zero at 
$u=0$. Then, $\Phi$ decreases with $u$ until it 
blows up at the singularity $r=0$.

Finally, let us see one way to obtain the Schwarzschild
black hole as the remnant of the scalar field collapse
described by the solutions in $S$. Each solution in
$S$ depends on two real parameters $M \geq 0$ and 
$B \leq 0$. From equation (\ref{20}) one can see that 
$B$ determines the scalar field strength. Therefore,
starting from Minkowski space-time $M=B=0$ at $u=0$
one turn the scalar field collapse on by increasing
$M$ and diminishing $B$. The scalar field strength
increases and a time-like naked singularity is formed
at $r=0$. Suppose now one fixes the value of $M=M_0$ 
and starts decreasing the scalar field strength by
making $B \to 0$. When $B=0$ the scalar field collapse
ends and we are left with a Schwarzschild black hole 
of mass $M_0$.

We would like to thank FAPEMIG for the invaluable 
financial support.

\begin{figure}
\begin{tabular}[c]{ccc}%
\mbox{\hspace{-0.2cm}} & (a) & (b) \\
\mbox{\hspace{-0.2cm}} & 
\resizebox{!}{5cm}{\includegraphics{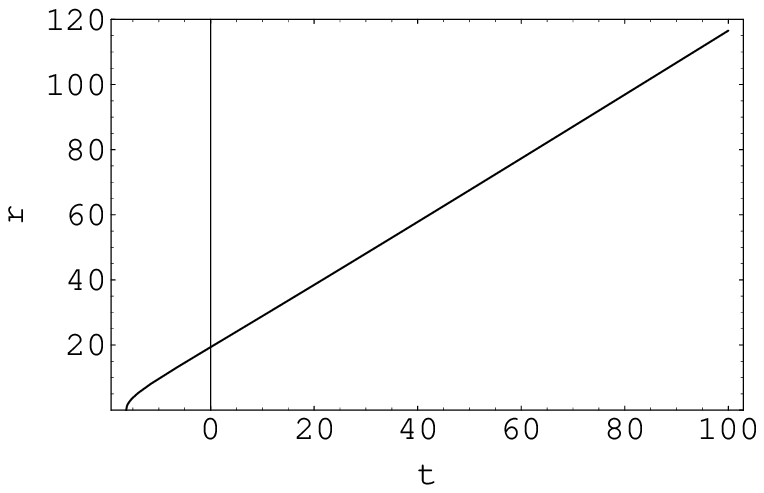}} &
\resizebox{!}{5cm}{\includegraphics{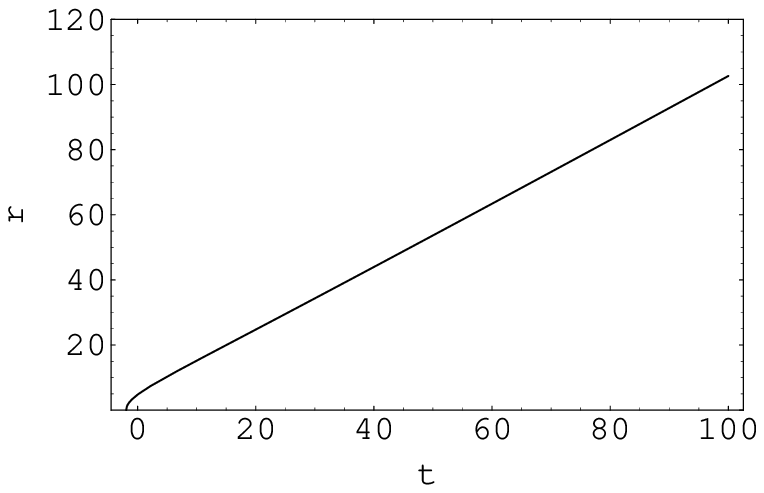}}%
\end{tabular}
\caption{Graphics of $r \times t$ for the space-time 
ST140 with two different values of $t_{\infty}$. (a) 
$t_{\infty} = 10^6$, (b) $t_{\infty} = 10^3$. }%
\end{figure}
\vskip.5in

\begin{figure}
\begin{tabular}[c]{ccc}%
\mbox{\hspace{-0.2cm}} & (a) & (b) \\
\mbox{\hspace{-0.2cm}} & 
\resizebox{!}{5cm}{\includegraphics{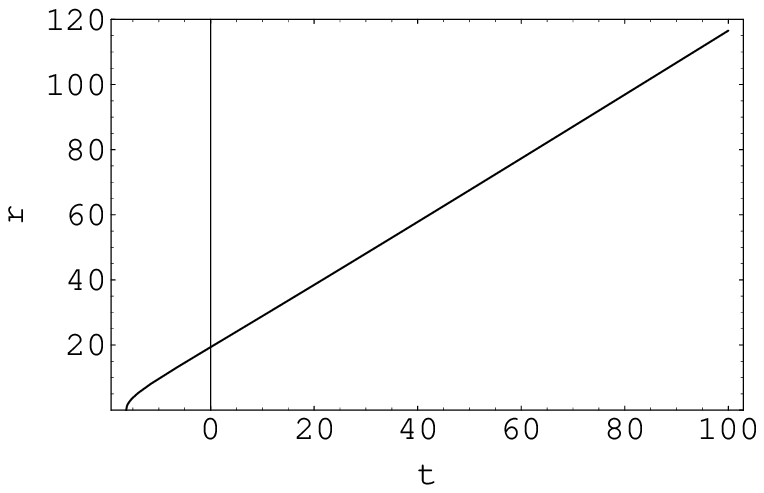}} &
\resizebox{!}{5cm}{\includegraphics{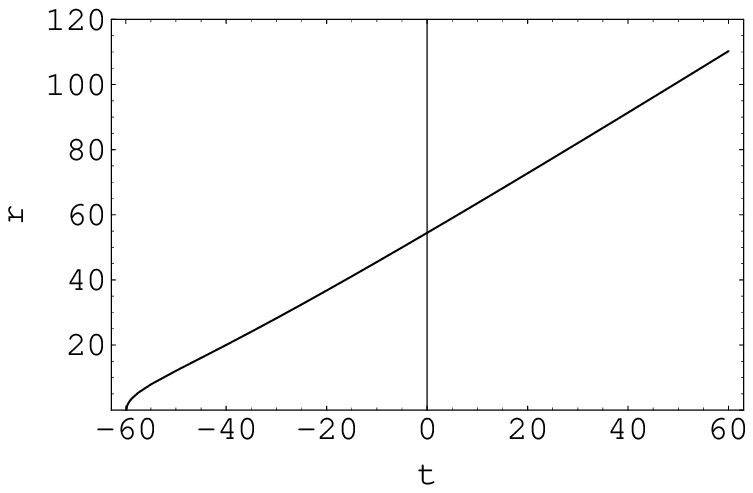}}%
\end{tabular}
\caption{Graphics of $r \times t$ for space-times with 
different values of $B \mbox{ and } M$. (a) ST140, 
(b) ST370. }%
\end{figure}


\begin{figure}
\begin{tabular}[c]{ccc}%
\mbox{\hspace{-0.2cm}} & (a) & (b) \\
\mbox{\hspace{-0.2cm}} & 
\resizebox{!}{5cm}{\includegraphics{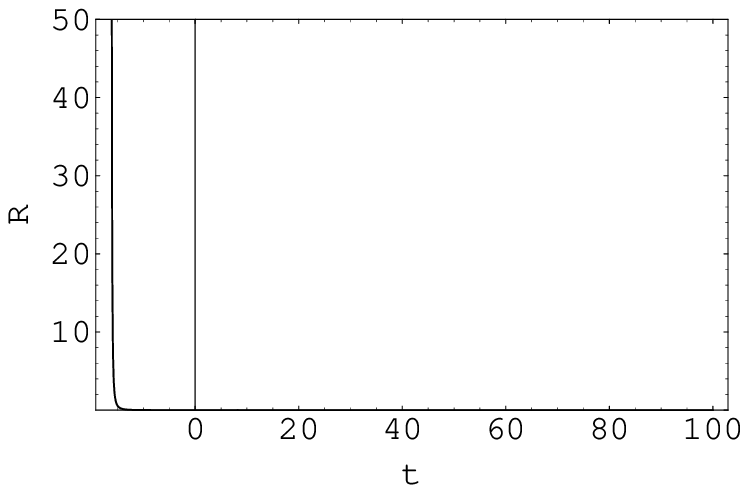}} &
\resizebox{!}{5cm}{\includegraphics{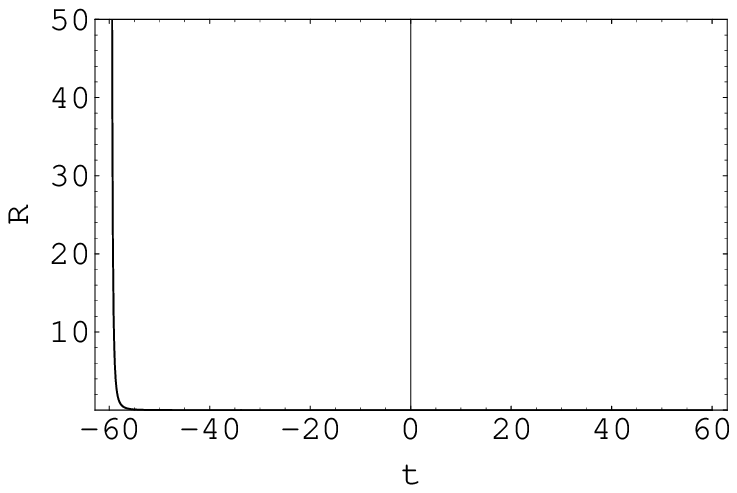}}%
\end{tabular}
\caption{Graphics of $R \times t$ for space-times with 
different values of $B \mbox{ and } M$. (a) ST140, 
(b) ST370. }%
\end{figure}
\vskip.5in

\begin{figure}
\begin{tabular}[c]{ccc}%
\mbox{\hspace{-0.2cm}} & (a) & (b) \\
\mbox{\hspace{-0.2cm}} & 
\resizebox{!}{5cm}{\includegraphics{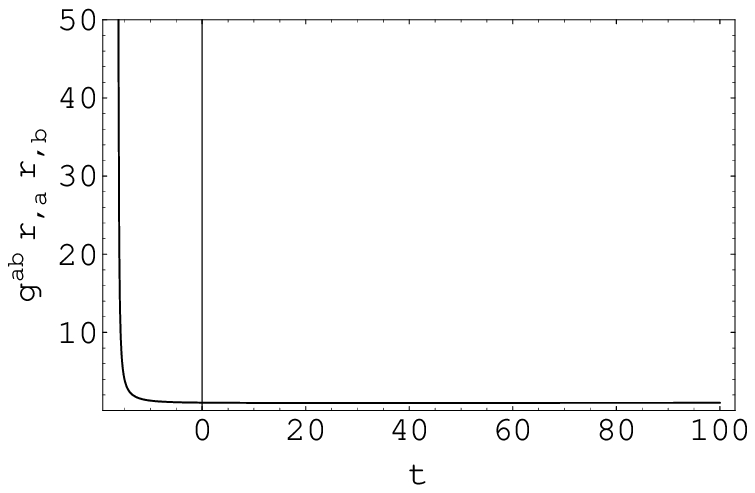}} &
\resizebox{!}{5cm}{\includegraphics{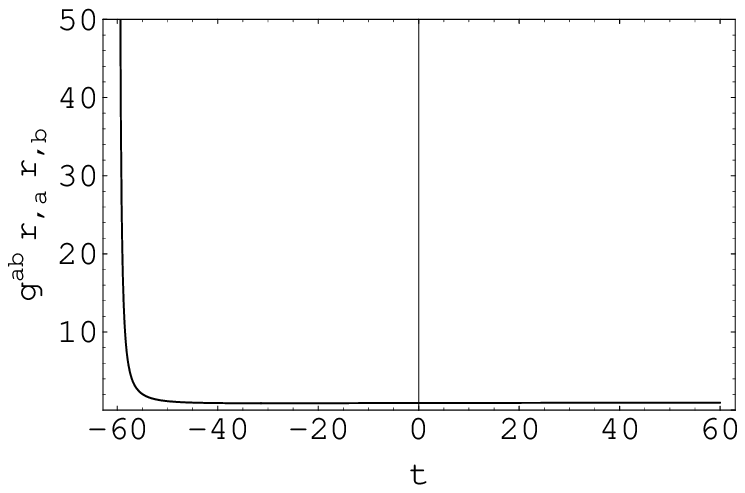}}%
\end{tabular}
\caption{Graphics of $g^{a b}r,_a r,_b \times \, t$ for 
space-times with different values of $B \mbox{ and } M$. 
(a) ST140, (b) ST370. }%
\end{figure}


\begin{figure}
\begin{tabular}[c]{ccc}%
\mbox{\hspace{-0.2cm}} & (a) & (b) \\
\mbox{\hspace{-0.2cm}} & 
\resizebox{!}{5cm}{\includegraphics{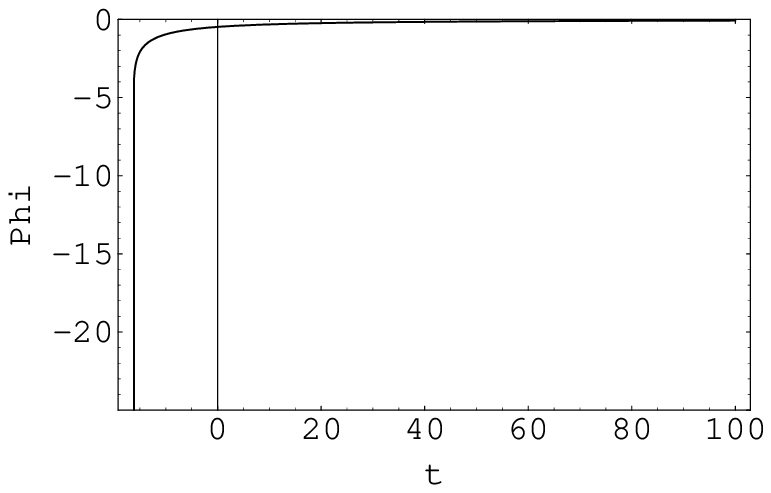}} &
\resizebox{!}{5cm}{\includegraphics{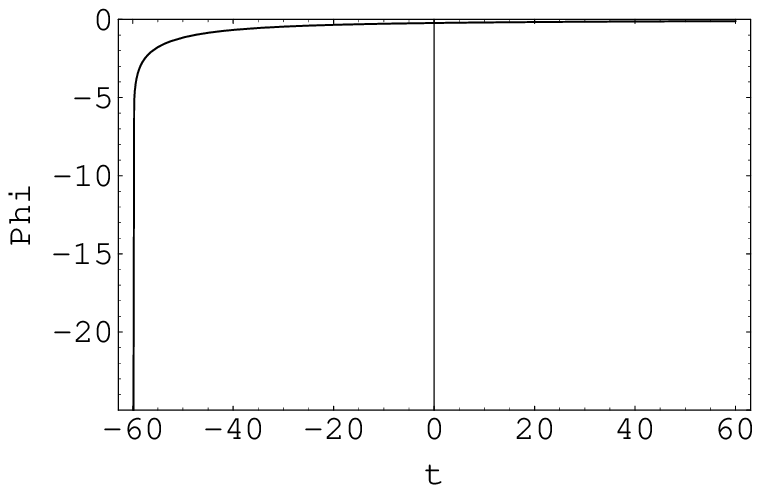}}%
\end{tabular}
\caption{Graphics of $\Phi \times t$ for space-times 
with different values of $B \mbox{ and } M$. (a) ST140, 
(b) ST370. }%
\end{figure}

\begin{figure}
\psfig{file=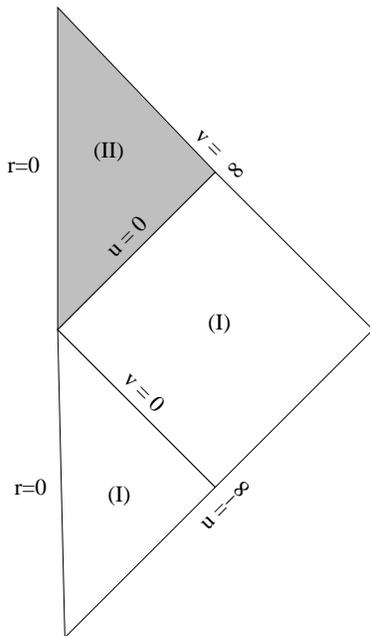}
\caption{Conformal diagram for a typical naked singularity 
solution. The null surface $u=0$ separates the
Minkowskian region (I) from the collapse region (II)  
where lies the time-like singularity at $r=0$.}
\end{figure}

\end{document}